# DSU-Net: An Attention-Enhanced Dense Skip U-Net for Breast Lesion Segmentation in Mammographic Images

Reza Bozorgpour *[1], Mohammadreza Soltany Sadrabadi[2]

[1]Department of Biomedical Engineering, University of Wisconsin-Milwaukee, Milwaukee, WI, USA
[2]Department of Mechanical Engineering, Northern Arizona University, Flagstaff, AZ, USA

*Corresponding author
E-mail: bozorgp2@uwm.edu

## Abstract

Breast cancer remains one of the leading causes of cancer-related mortality among women worldwide, making early detection essential for effective treatment. Mammography is the primary screening modality; however, accurate delineation of suspicious lesions remains challenging and subject to inter-observer variability. Automated segmentation methods can assist radiologists by providing consistent and efficient lesion localization.

This study presents DSU-Net, an attention-enhanced Dense Skip U-Net architecture for automated breast lesion segmentation in mammographic images. The proposed framework integrates dense skip connections and attention mechanisms to improve feature propagation, preserve spatial information, and enhance lesion boundary delineation. Experiments were conducted using the Curated Breast Imaging Subset of the Digital Database for Screening Mammography (CBIS-DDSM). To address severe foreground-background imbalance, a composite loss function combining Dice loss, focal loss, and binary cross-entropy loss was employed during training.

The proposed model achieved a Dice Similarity Coefficient of 0.9421, an Intersection over Union of 0.8905, an accuracy of 0.9711, and an AUC-ROC of 0.9878 on the validation dataset. Qualitative evaluation demonstrated accurate delineation of lesions with varying sizes and morphologies, while quantitative results confirmed robust discrimination between lesion and background regions.

These findings demonstrate that DSU-Net provides accurate and reliable breast lesion segmentation in mammographic images and highlights the potential of attention-guided deep learning for computer-aided breast cancer screening and diagnosis.

# 1. Introduction

Breast cancer remains one of the leading health concerns affecting women and is the most frequently diagnosed cancer among females in the United States [1, 2]. Improving patient outcomes largely depends on the early detection of malignant breast lesions, as diagnosis at an earlier stage can significantly reduce disease progression and mortality. For this reason, screening programs based on breast self-examination and mammography have become widely adopted in clinical practice [2, 3].

Although mammography is a cornerstone of breast cancer screening, its diagnostic performance can be limited in women with dense breast tissue. To overcome this limitation, ultrasound (US) is often used as a complementary imaging modality, leading to improved detection and characterization of breast abnormalities [4, 5]. In addition to enhancing diagnostic accuracy, US offers several practical advantages, including real-time imaging, low cost, and a non-invasive examination process, making it an attractive tool for routine breast assessment [6-8].

Despite these advantages, analyzing breast ultrasound images remains challenging. The identification and delineation of breast masses require considerable expertise and are often influenced by operator experience, making the process labor-intensive and time-consuming for radiologists. These challenges have motivated extensive research into computer-aided diagnosis systems and machine learning techniques aimed at supporting clinicians, improving workflow efficiency, and reducing the burden of image interpretation [9].

Recent advances in deep learning have further accelerated the development of intelligent medical imaging systems. By learning complex image patterns directly from data, deep learning models have demonstrated considerable potential for assisting clinical decision-making and extending expert-level support to settings where specialized resources are limited [10, 11]. In the context of breast ultrasound imaging, automated lesion segmentation represents a particularly important application. Accurate deep learning-based segmentation can streamline image analysis, reduce manual effort, and provide more consistent lesion characterization, thereby helping to improve both efficiency and diagnostic decision-making in clinical practice [12-16].

Breast ultrasound segmentation has evolved from traditional image-analysis and machine learning approaches toward deep learning-based solutions. Although earlier methods were explored extensively for automated lesion detection and segmentation, their ability to capture complex lesion characteristics remained limited [17-19]. Deep learning models overcome this limitation by learning representative features directly from imaging data, leading to their widespread adoption in breast cancer imaging applications [19-26]. Architectures ranging from CNNs to attention-guided and encoder–decoder frameworks have demonstrated improved lesion delineation and can assist clinicians by streamlining the diagnostic process [21, 27].

U-Net remains the most widely adopted deep learning architecture for breast ultrasound lesion segmentation because of its encoder–decoder design and strong feature-learning capability [28]. As a result, numerous studies have focused on enhancing the original U-Net to improve segmentation accuracy and robustness [29-35]. Several modifications have been proposed to address the limitations of conventional U-Net. Byra et al. incorporated selective kernels and attention mechanisms to adapt receptive fields dynamically [29], while Kumar et al. developed a multi-U-Net framework for automatic real-time segmentation without requiring an initial seed point [32]. Guo et al. introduced dropout layers to reduce overfitting while preserving lesion boundaries [31], and Punn et al. proposed RCA-IUnet, which improved segmentation across different lesion sizes but at the cost of increased computational complexity [36].

Attention-based approaches have also shown promising results. Yan et al. combined hybrid dilated convolutions with a lesion-focused loss function to improve segmentation performance [37], whereas Tong et al. employed a mixed-attention loss strategy in a modified U-Net architecture [38]. To better capture lesion variability, a multi-scale fusion U-Net was later developed with a focal loss function to address class imbalance [39]. However, many of these studies were validated on limited datasets, making it difficult to assess their generalizability [29, 32, 39]. To address this issue, a saliency-guided morphology-aware U-Net was evaluated across five datasets [40]. More recently, Chen et al. replaced conventional convolutional layers with channel-attention and self-attention modules to improve feature extraction at multiple scales. Despite these advances, false-positive predictions remain a challenge.
In addition to U-Net variants, alternative architectures such as transfer learning-based FCN-AlexNet [2] and boundary-guided region-aware networks [7] have achieved strong segmentation performance. Nevertheless, their complex designs require substantial computational resources, leading to longer inference times. Moreover, performance degradation on unseen datasets and the relatively small datasets used for validation raise concerns about their robustness and generalizability.

# 2. Motivation

Breast lesion segmentation in mammographic images remains challenging due to large variations in lesion morphology, low contrast between lesions and surrounding tissue, and the severe class imbalance between lesion and background regions. Although U-Net and its variants have achieved promising results, many existing approaches struggle to effectively capture both local lesion details and high-level contextual information, particularly when lesion boundaries are indistinct or lesion sizes vary substantially.
Another limitation of conventional encoder-decoder architectures is the progressive loss of information during feature extraction. While deeper networks can learn more discriminative representations, they may also suffer from reduced feature reuse and weaker gradient propagation. Furthermore, not all features transferred through skip connections contribute equally to lesion segmentation, potentially introducing irrelevant background information into the decoder pathway.
To address these challenges, this study proposes a Dense and Spatial Attention U-Net (DSU-Net) architecture that integrates dense feature propagation within the bottleneck stage and attention-guided skip connections throughout the decoder. Dense connectivity promotes feature reuse and facilitates information flow across layers, while attention mechanisms enable the network to selectively emphasize lesion-relevant features during reconstruction of the segmentation map. In addition, a hybrid loss function combining Dice, Focal, and Binary Cross-Entropy losses is employed to improve segmentation performance under severe class imbalance conditions.
A patient-level training and validation strategy is further adopted to eliminate information leakage between datasets and provide a more reliable assessment of model generalization. Through the integration of dense feature learning, attention-guided feature selection, and hybrid loss optimization, the proposed framework aims to improve the accuracy and robustness of breast lesion segmentation in mammographic images.

# 3. Materials and Methods

## 3.1. Dataset

The proposed segmentation framework was developed using the Curated Breast Imaging Subset of the Digital Database for Screening Mammography (CBIS-DDSM), a publicly available and extensively used benchmark dataset for breast cancer research. The dataset contains digitized mammographic images together with expert annotations describing lesion locations, pathology information, assessment scores, and corresponding segmentation masks. Both mass and calcification cases from the CBIS-DDSM training dataset were utilized in this study. Metadata associated with each lesion were obtained from the *mass_case_description_train_set.csv* and *calc_case_description_train_set.csv*. These files contain patient identifiers, pathology labels, Breast Imaging Reporting and Data System (BI-RADS) assessment values, cropped image locations, and ROI mask locations. To construct the segmentation dataset, the image paths provided in the metadata files were automatically parsed and matched with the corresponding JPEG image directories. For each case, the cropped lesion image and its associated ROI mask were identified and verified. Cases for which both the cropped mammogram and the corresponding segmentation mask were successfully located were retained. A master dataset was subsequently generated containing the image path, mask path, pathology label, assessment score, and patient identifier for each valid sample. A master dataset was subsequently generated containing the image path, mask path, pathology label, assessment score, and patient identifier for each valid sample.

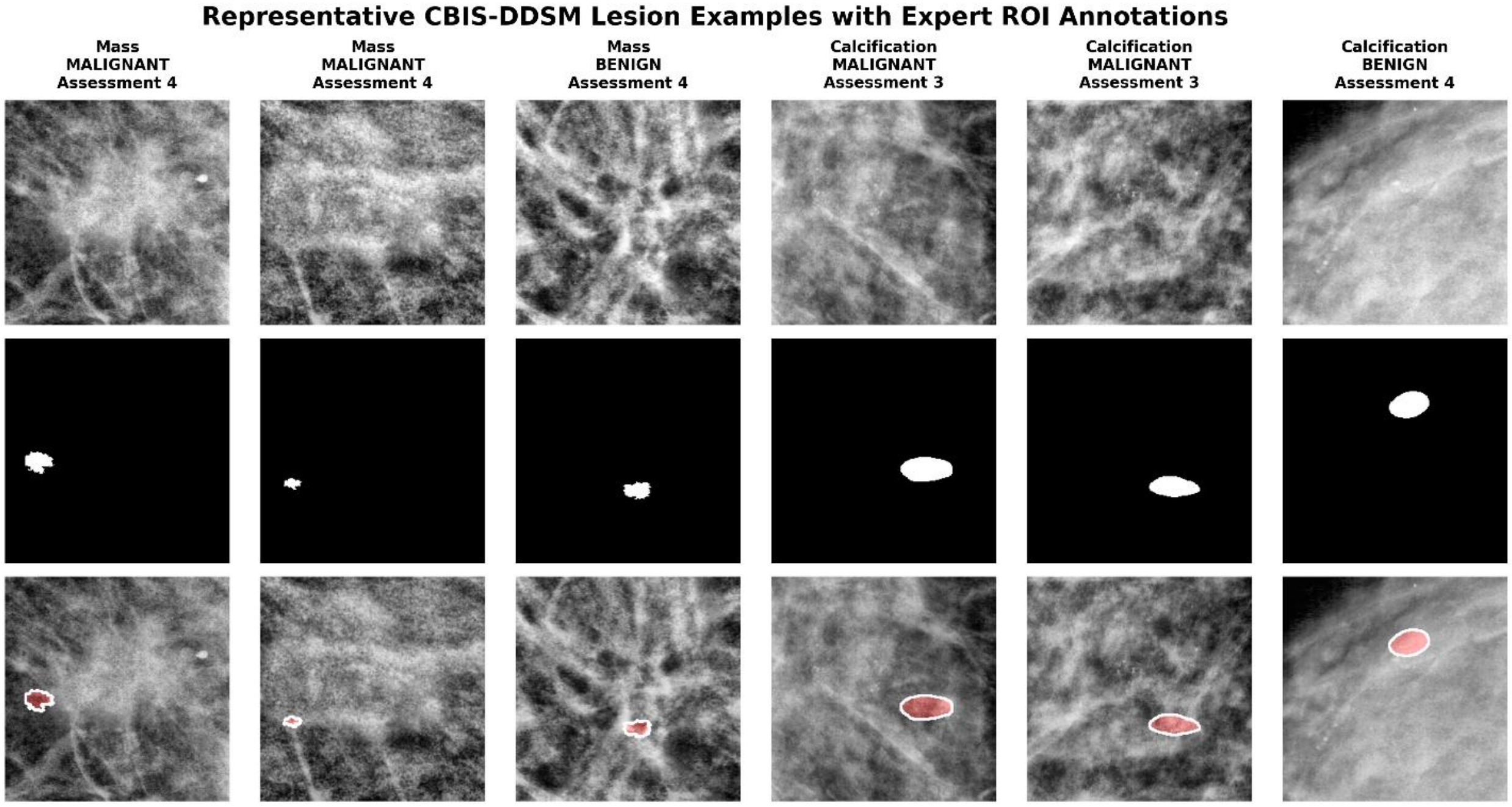

Figure 1: Representative CBIS-DDSM mammogram ROIs with corresponding expert segmentation masks and lesion overlays.

## 3.2. Image Preprocessing

The CBIS-DDSM images and corresponding ROI masks underwent a standardized preprocessing procedure prior to training. Since mammographic images within the dataset vary in size, format, and intensity distribution, a sequence of preprocessing operations was applied to ensure consistency across all samples. The preprocessing pipeline included image loading, resizing, color-space conversion, mask binarization, normalization, and tensor conversion. These steps were designed to generate uniform inputs suitable for deep learning-based segmentation while preserving lesion characteristics and anatomical details. The complete preprocessing workflow is summarized in *Table 1*.

Table 1: Preprocessing pipelines applied to mammographic images and ROI masks.

| Step | Description | Parameter/Setting |
|---|---|---|
| **Image Loading** | Cropped lesion images loaded using OpenCV | 3-channel color images |
| **Mask Loading** | ROI masks loaded as grayscale images | Single-channel grayscale |
| **Image Resizing** | Standardization of image dimensions | 256 × 256 pixels |
| **Mask Resizing** | Standardization of mask dimensions | 256 × 256 pixels |
| **Color-Space Conversion** | Conversion to RGB format | BGR → RGB |
| **Mask Binarization** | Generation of binary segmentation masks | Threshold = 127 |
| **Mask Normalization** | Scaling mask values | Range [0,1] |
| **Image Normalization** | Intensity normalization using predefined statistics | Mean = (0.485, 0.456, 0.406) |
| **Tensor Conversion** | Conversion of images and masks into PyTorch tensors | ToTensorV2 |
| | | Std = (0.229, 0.224, 0.225) |

All cropped lesion images were resized to 256 × 256 pixels to provide a fixed input size for the network while reducing computational cost. ROI masks were converted into binary segmentation maps through thresholding and subsequently normalized to the range [0,1]. Image intensities were normalized using standard mean and standard deviation values commonly adopted in convolutional neural network applications. Finally, the processed images and masks were converted into PyTorch tensors, enabling efficient GPU-based training and inference.

### 3.3. Data Augmentation

Breast lesion datasets are often limited in size and may not adequately represent the wide variability encountered in clinical practice. To improve model generalization and reduce overfitting, an extensive data augmentation strategy was applied to the training images. The augmentations were implemented using the Albumentations library and were performed online during training. Validation images were excluded from augmentation to ensure an unbiased assessment of model performance. The augmentation techniques and their corresponding parameter settings are summarized in *Table 2*.

Table 2: Data augmentation techniques applied during training.

| Augmentation | Parameter |
|---|---|
| **Horizontal Flip** | Probability = 0.5 |
| **Vertical Flip** | Probability = 0.3 |
| **Random Rotation** | ±20° |
| **Shift-Scale-Rotate** | Shift = 10%, Scale = 20%, Rotation = ±20° |
| **Brightness Adjustment** | ±30% |
| **Contrast Adjustment** | ±30% |
| **Elastic Transformation** | $\alpha = 1$, $\sigma = 50$ |
| **Grid Distortion** | Probability = 0.3 |
| **Gaussian Noise** | Variance = 10–50 |
| **Normalization** | Mean = (0.485, 0.456, 0.406), Std = (0.229, 0.224, 0.225) |

The augmentation pipeline incorporated both geometric and intensity-based transformations. Geometric operations such as flipping, rotation, scaling, elastic deformation, and grid distortion exposed the network to variations in lesion shape, size, and orientation. Intensity-based transformations, including brightness and contrast adjustments and Gaussian noise injection, improved robustness to imaging variations and acquisition conditions. Collectively, these augmentations increased the diversity of the training set and enhanced the model's ability to generalize to previously unseen mammographic images.

### 3.4. Patient-Level Data Splitting

To prevent information leakage between training and validation sets, data partitioning was performed at the patient level rather than the image level. Unique patient identifiers were extracted from the master dataset and randomly divided into training and validation groups using an 80:20 split. All images belonging to a given patient were assigned exclusively to one subset. This strategy ensured that lesions from the same patient could not appear in both training and validation datasets.

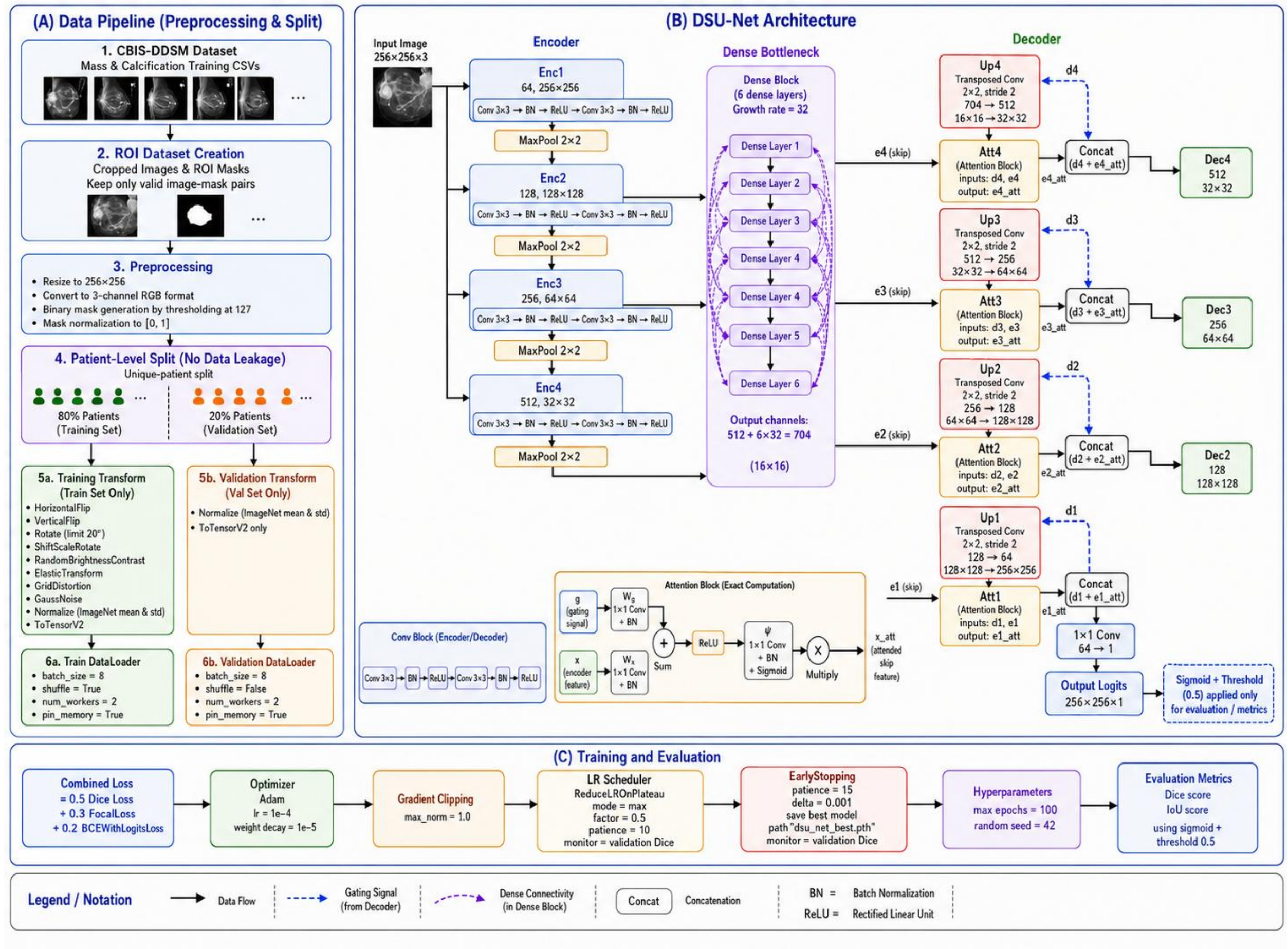

Figure 2: Training pipeline and DSU-Net architecture for breast lesion segmentation.

### 3.5. DSU-Net Architecture

To accurately segment breast lesions from mammographic images, a Dense and Spatial Attention U-Net (DSU-Net) architecture was developed. The proposed network combines the encoder-decoder structure of U-Net with DenseNet-inspired feature propagation and attention-guided skip connections. The objective of this design is to improve feature reuse, enhance gradient flow during training, and enable the network to focus on diagnostically relevant regions while suppressing background information.

The overall architecture consists of an encoder path, a dense bottleneck module, and an attention-guided decoder path. The network receives a $256 \times 256$ RGB image as input and generates a single-channel segmentation map representing the predicted lesion region.

#### 3.5.1.Encoder Path

The encoder is composed of four hierarchical feature extraction stages. Each stage contains two consecutive $3 \times 3$ convolutional layers followed by batch normalization and ReLU activation functions. The number of feature channels increases progressively through the network, allowing the extraction of increasingly abstract image representations.

The first encoder block generates 64 feature maps from the input image. Subsequent encoder blocks produce 128, 256, and 512 feature maps, respectively. Between adjacent encoder stages, max-pooling operations with a kernel size of $2 \times 2$ are applied to reduce spatial resolution and increase the receptive field. This hierarchical downsampling enables the network to capture both local texture information and larger contextual structures associated with breast lesions.

#### 3.5.2.Dense Bottleneck Module

At the deepest level of the network, a Dense Block is incorporated to strengthen feature propagation and encourage feature reuse. Unlike conventional convolutional bottlenecks, the Dense Block establishes direct connections between all preceding and subsequent layers.

The bottleneck consists of six densely connected layers with a growth rate of 32. Each layer receives the concatenated outputs of all previous layers and performs feature extraction through a sequence of batch normalization, ReLU activation, a $1 \times 1$ bottleneck convolution, and a $3 \times 3$ convolution.

By concatenating features from multiple depths, the Dense Block improves information flow throughout the network and mitigates the vanishing-gradient problem commonly encountered in deep architectures. Furthermore, dense connectivity allows the model to exploit both low-level and high-level image features simultaneously, which is particularly beneficial for segmenting lesions with varying sizes, shapes, and texture characteristics.

#### 3.5.3.Attention-Guided Decoder

The decoder path reconstructs the segmentation map by progressively recovering spatial resolution through four upsampling stages. Each stage begins with a transposed convolution layer that doubles the spatial dimensions of the feature maps.

To improve lesion localization, attention gates are incorporated into all skip connections between encoder and decoder stages. These attention modules receive feature information from both pathways and compute attention coefficients that selectively emphasize lesion-related structures while suppressing irrelevant background responses.

For a given decoder stage, the upsampled decoder features are first used as gating signals. These signals are combined with the corresponding encoder features through learnable $1 \times 1$ convolutional transformations and batch normalization layers. The resulting attention map is passed through a sigmoid activation function and used to weigh the encoder features before concatenation with the decoder features.

The attended encoder features are then merged with the decoder features and processed using convolutional refinement blocks consisting of two $3 \times 3$ convolutional layers, batch normalization, and ReLU activation.

This attention-guided fusion mechanism allows the network to focus on clinically relevant lesion boundaries and improves segmentation precision.

#### 3.5.4.Output Layer

Following the final decoder stage, a $1 \times 1$ convolutional layer is applied to generate a single-channel output map. This layer transforms the learned feature representation into pixel-wise lesion predictions.

The network produces raw logits during training, which are subsequently processed by the loss function. During evaluation, a sigmoid activation function is applied to convert logits into probability values ranging from 0 to 1. Pixels with probabilities greater than 0.5 are classified as lesion pixels, while the remaining pixels are assigned to the background class.

### 3.6. Loss Function Design

Breast lesion segmentation is inherently challenging due to the severe class imbalance between lesion and background pixels. In mammographic images, lesion regions typically occupy only a small fraction of the image area, causing conventional loss functions to become biased toward the dominant background class. To address this limitation, a hybrid loss function combining Dice loss, Focal loss, and Binary Cross-Entropy (BCE) loss was adopted. By integrating region-based and pixel-based objectives, the proposed loss function aims to improve lesion localization, boundary delineation, and overall segmentation accuracy.

The total loss function is defined as

$$L_{\text{Total}} = 0.5L_{\text{Dice}} + 0.3L_{\text{Focal}} + 0.2L_{\text{BCE}} \quad \text{Eq. 1}$$

Where $L_{\text{Dice}}$, $L_{\text{Focal}}$, and $L_{\text{BCE}}$ denote the Dice loss, Focal loss, and Binary Cross-Entropy loss, respectively. The weighting coefficients were selected to balance overlap-based optimization, hard-sample learning, and pixel-wise classification performance. Dice loss was incorporated to directly maximize the overlap between the predicted segmentation and the ground-truth lesion mask and is defined as

$$L_{\text{Dice}} = 1 - \frac{2\sum_{i=1}^{N} p_i\, g_i + \varepsilon}{\sum_{i=1}^{N} p_i + \sum_{i=1}^{N} g_i + \varepsilon} \quad \text{Eq. 2}$$

where $p_i$and $g_i$represent the predicted probability and ground-truth label for pixel $i$, respectively, $N$denotes the total number of pixels, and $\varepsilon$is a small smoothing constant introduced to avoid numerical instability. Because Dice loss directly optimizes the overlap between predicted and reference lesion regions, it is particularly effective for highly imbalanced segmentation problems. To provide stable pixel-level supervision during optimization, Binary Cross-Entropy loss was also incorporated and is expressed as

$$L_{\text{BCE}} = -\frac{1}{N}\sum_{i=1}^{N}[g_i \ln(p_i) + (1 - g_i)\ln(1 - p_i)] \quad \text{Eq. 3}$$

Although BCE loss contributes to reliable convergence, it may still be influenced by the large number of background pixels present in mammographic images. Therefore, Focal loss was included to emphasize difficult-to-classify pixels and reduce the contribution of easily classified background regions. The Focal loss is given by

$$L_{\text{Focal}} = -\frac{1}{N}\sum_{i=1}^{N} \alpha(1 - p_{t,i})^{\gamma} \ln(p_{t,i}) \quad \text{Eq. 4}$$

Where

$$p_{t,i} = \begin{cases} p_i, & \text{if } g_i = 1 \\ 1 - p_i, & \text{if } g_i = 0 \end{cases} \quad \text{Eq. 5}$$

In this formulation, $\alpha$and $\gamma$denote the balancing and focusing parameters, respectively. Following the implementation adopted in this study, these parameters were set to $\alpha = 1$and $\gamma = 2$. The focal loss mechanism assigns greater emphasis to difficult samples during training while reducing the influence of correctly classified pixels. The combination of Dice, Focal, and BCE losses enables the proposed DSU-Net to simultaneously maximize lesion-region overlap, improve boundary delineation, and maintain stable pixel-level optimization. By exploiting the complementary strengths of these loss functions, the network achieves more robust and accurate breast lesion segmentation across lesions exhibiting substantial variability in size, shape, and appearance.

### 3.7. Model Training

The DSU-Net model was implemented using the PyTorch deep learning framework and trained using supervised learning. Model optimization was performed using the Adam optimizer with an initial learning rate of $1 \times 10^{-4}$ and a weight decay coefficient of $1 \times 10^{-5}$. Training was conducted using a batch size of 8 and a maximum of 100 epochs. The complete training configuration and optimization parameters used in this study are summarized in Table *3*.

Table 3: DSU-Net training parameters.

| Parameter | Value |
|---|---|
| **Input Image Size** | 256 × 256 |
| **Batch Size** | 8 |
| **Number of Epochs** | 100 |
| **Initial Learning Rate** | $1 \times 10^{-4}$ |
| **Optimizer** | Adam |
| **Weight Decay** | $1 \times 10^{-5}$ |
| **Learning Rate Scheduler** | ReduceLROnPlateau |
| **LR Reduction Factor** | 0.5 |
| **Scheduler Patience** | 10 epochs |
| **Early Stopping Patience** | 15 epochs |
| **Early Stopping Delta** | 0.001 |
| **Gradient Clipping** | Max norm = 1.0 |
| **Loss Function** | 0.5 Dice + 0.3 Focal + 0.2 BCE |
| **Validation Split** | 20% |
| **Random Seed** | 42 |
| **Output Activation (Evaluation)** | Sigmoid |
| **Segmentation Threshold** | 0.5 |
| **Evaluation Metrics** | Dice Similarity Coefficient (DSC), Intersection over Union (IoU) |

To improve training stability, gradient clipping was applied after backpropagation with a maximum gradient norm of 1.0. This strategy prevents excessively large parameter updates and reduces the likelihood of unstable optimization behavior.

A ReduceLROnPlateau learning-rate scheduler was employed to automatically adapt the learning rate during training. The scheduler monitored the validation Dice score and reduced the learning rate by a factor of 0.5 whenever validation performance failed to improve for ten consecutive epochs. This adaptive learning-rate strategy enabled more efficient convergence and improved optimization of the segmentation network.

### 3.8. Early Stopping and Model Selection

To prevent overfitting and identify the most effective model, an early stopping strategy was incorporated during training. After each epoch, the validation Dice score was evaluated and compared with the best performance achieved thus far.

Model parameters were automatically saved whenever an improvement in validation Dice score was observed. Training was terminated when the validation Dice score failed to improve by at least 0.001 for 15 consecutive epochs. This approach prevented unnecessary training after convergence and reduced the risk of overfitting to the training data.

Upon completion of training, the model corresponding to the highest validation Dice score was reloaded and designated as the final segmentation model used for evaluation.

## 4. Results

The proposed DSU-Net model demonstrated stable convergence during training on the CBIS-DDSM dataset. The training configuration and hyperparameters used in this study are summarized in *Table 4*. Although training was configured for a maximum of 100 epochs, early stopping terminated the process after 24 epochs. The best validation performance, presented in *Table 5*, was achieved at Epoch 9, yielding a Dice score of 0.9442 and an IoU score of 0.9065. These results indicate strong agreement between the predicted lesion regions and the expert-annotated ground-truth masks.

Table 4: Training configuration used for DSU-Net training

| Parameter | Value |
|---|---|
| **Architecture** | DSU-Net |
| **Input Size** | 256 × 256 |
| **Batch Size** | 8 |
| **Learning Rate** | $1 \times 10^{-4}$ |
| **Optimizer** | Adam |
| **Weight Decay** | $1 \times 10^{-5}$ |
| **Loss Function** | Combined Loss (Dice + Focal + BCE) |
| **Early Stopping Patience** | 15 |
| **Maximum Epochs** | 100 |
| **Actual Epochs Trained** | 24 |

Table 5: Best validation performance achieved by DSU-Net.

| Metric | Value |
|---|---|
| **Validation Loss** | 0.1266 |
| **Dice Score** | 0.9442 |
| **IoU Score** | 0.9065 |
| **Best Epoch** | 9 |

The quantitative results presented in *Table 5* demonstrate that DSU-Net achieved high segmentation accuracy on the validation dataset. To further evaluate the quality of the predicted segmentations, representative qualitative results are shown in Figure 3. The selected examples include lesions with substantially different sizes and morphologies, ranging from large irregular masses to small, localized lesions occupying only a small fraction of the image.

As illustrated in Figure 3, the predicted boundaries closely match the expert annotations across all representative cases. For larger lesions, DSU-Net accurately captured the overall lesion extent and complex boundary geometry, while for smaller lesions the model successfully localized the lesion region despite the pronounced class imbalance between lesion and background pixels. The predicted contours preserved the shape and spatial location of the annotated lesions with only minor deviations along highly irregular boundary regions. The strong agreement between the predicted and reference contours is consistent with the Dice score of 0.9442 and IoU score of 0.9065 reported in *Table 5*, confirming the effectiveness of the proposed DSU-Net architecture for mammographic lesion segmentation.

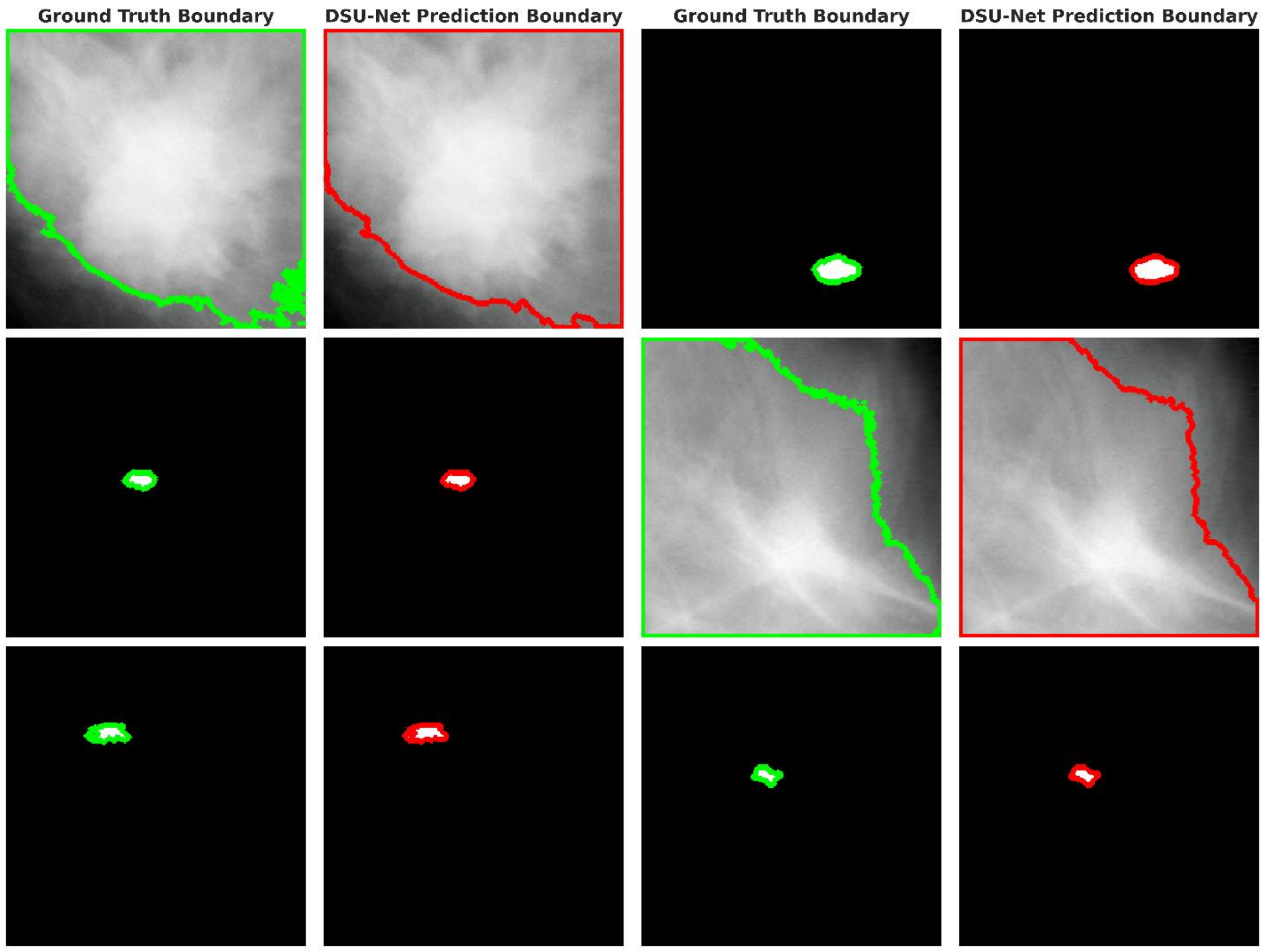

Figure 3: Comparison of expert-annotated lesion boundaries (green) and DSU-Net predicted boundaries (red) for representative validation samples.

While Figure 3 demonstrates that DSU-Net accurately reproduces lesion boundaries across representative validation samples, visual inspection alone does not fully characterize segmentation performance. In particular, qualitative comparisons do not quantify the balance between lesion detection and false-positive predictions. Therefore, additional evaluation metrics were computed to provide a more comprehensive assessment of model behavior. The resulting performance measures are summarized in Table 6.

Table 6: Comprehensive segmentation performance metrics of DSU-Net on the validation dataset.

| Metric | Value |
|---|---|
| **Dice Score** | 0.9421 |
| **IoU Score** | 0.8905 |
| **Accuracy** | 0.9711 |
| **Precision** | 0.8971 |
| **Recall (Sensitivity)** | 0.9918 |
| **Specificity** | 0.9647 |
| **AUC-ROC** | 0.9878 |

As shown in *Table 6*, DSU-Net achieved strong performance across all evaluation metrics. The model obtained a Dice score of 0.9421 and an IoU score of 0.8905, indicating substantial overlap between the predicted and ground-truth lesion regions. The high recall (0.9918) demonstrates that the model successfully identified nearly all lesion pixels, while the precision of 0.8971 indicates a relatively low number of false-positive predictions. Furthermore, the AUC-ROC of 0.9878 confirms excellent discrimination between lesion and background pixels. Collectively, these results demonstrate that the proposed DSU-Net architecture provides accurate and robust breast lesion segmentation on the CBIS-DDSM dataset. A visual summary of the segmentation performance metrics is provided in Figure 4.

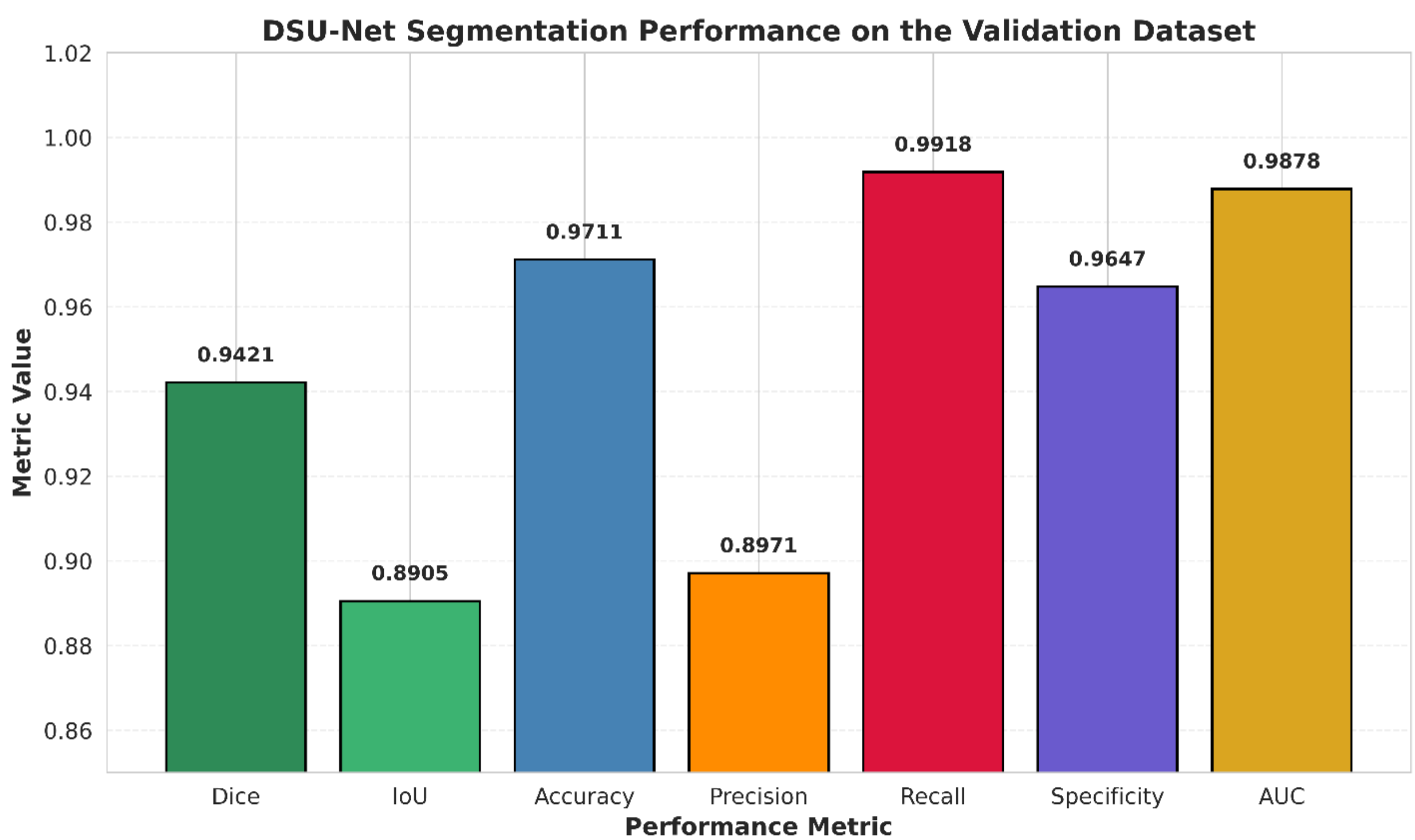

Figure 4: Segmentation performance metrics achieved by DSU-Net on the CBIS-DDSM validation dataset.

## 5. Discussion

The results demonstrate that the proposed DSU-Net architecture is capable of accurately segmenting breast lesions from mammographic regions of interest despite the substantial class imbalance inherent to mammography datasets. In the CBIS-DDSM dataset, lesion pixels typically occupy only a small portion of the image while the majority of pixels belong to the background. This imbalance presents a significant challenge for segmentation networks because conventional optimization strategies often favor the dominant background class. The high Dice score (0.9442) and IoU score (0.9065) obtained by DSU-Net indicate that the network successfully learned to localize lesion regions while maintaining strong overlap with expert-annotated ground-truth masks.

The qualitative results further support quantitative findings. As demonstrated in Figure 3, the predicted lesion boundaries closely follow the expert annotations across lesions exhibiting substantially different sizes and morphologies. The model accurately captured both large irregular masses and small localized lesions, suggesting that the learned feature representations remain effective across a wide range of lesion scales. This behavior is particularly important in mammographic analysis because lesion appearance can vary considerably due to differences in tissue density, lesion shape, and image contrast. The ability of DSU-Net to preserve lesion morphology while maintaining accurate localization indicates that the network effectively combines global contextual information with fine-grained boundary features.

The additional performance metrics reported in *Table 6* provide further insight into the behavior of the model. The recall of 0.9918 indicates that nearly all lesion pixels were successfully identified, which is a desirable property in computer-aided breast cancer detection systems where missed lesions may have significant clinical consequences. Although maximizing sensitivity can sometimes lead to increased false-positive predictions, the model simultaneously achieved a precision of 0.8971 and a specificity of 0.9647, demonstrating that the high sensitivity was not obtained at the expense of excessive over-segmentation. The balance between recall, precision, and specificity suggests that the combined loss formulation used during training effectively constrained both false-negative and false-positive errors.

The training strategy also contributed to the observed performance. The use of a composite loss function consisting of Dice loss, Focal loss, and Binary Cross-Entropy loss allowed the network to address multiple optimization objectives simultaneously. Dice loss directly optimizes region overlap, Focal loss emphasizes difficult and underrepresented pixels, and Binary Cross-Entropy provides stable pixel-wise supervision. This combination is particularly advantageous for mammographic lesion segmentation because lesion boundaries often exhibit low contrast and irregular shapes. The resulting performance indicates that the complementary strengths of these loss functions contributed to improved lesion delineation and robust convergence during training.

Another notable observation is the stability of the optimization process. Although training was configured for 100 epochs, early stopping terminated the process after only 24 epochs, with the best validation performance achieved at Epoch 9. This behavior suggests that the network rapidly converged to a high-quality solution and that prolonged training was unnecessary. The absence of performance degradation after the optimal epoch indicates that the regularization strategy, including weight decay and early stopping, was effective in reducing overfitting despite the limited size of the medical imaging dataset.

The AUC-ROC value of 0.9878 further demonstrates the strong discriminative capability of the proposed architecture. This result indicates that the network effectively separates lesion and background pixels across a wide range of decision thresholds rather than relying on a single segmentation threshold. Combined with the high accuracy of 0.9711, these findings suggest that DSU-Net learned robust feature representations capable of generalizing across the diverse lesion appearances present in the CBIS-DDSM dataset.

Overall, the quantitative and qualitative results collectively demonstrate that DSU-Net provides accurate and reliable breast lesion segmentation in mammographic images. The strong overlap metrics, high sensitivity, excellent discriminative performance, and accurate preservation of lesion morphology indicate that the proposed architecture successfully addresses several key challenges associated with mammographic image analysis, including severe class imbalance, lesion heterogeneity, and boundary ambiguity. These findings support the potential applicability of DSU-Net as a component of computer-aided diagnosis systems for breast cancer detection and assessment.

# 6. Limitation

Several limitations of the present study should be acknowledged. First, model development and evaluation were performed using the CBIS-DDSM dataset, which, although widely used for breast imaging research, may not fully represent the diversity of mammographic images encountered in clinical practice. Variations in imaging protocols, acquisition devices, patient populations, and lesion characteristics across institutions may affect model generalizability. Consequently, external validation on independent multi-center datasets is necessary to further assess the robustness of the proposed approach.

Second, the study focused exclusively on lesion segmentation and did not investigate downstream clinical tasks such as lesion classification, malignancy prediction, or breast cancer risk assessment. While accurate segmentation provides a critical foundation for these applications, additional studies are required to determine how the generated lesion masks influence subsequent diagnostic analyses.

Third, all images were resized to a fixed spatial resolution prior to training to maintain computational efficiency and standardized model inputs. Although this approach facilitates network optimization, it may reduce fine-scale image details that could be relevant for the segmentation of very small lesions or subtle lesion boundaries.

Finally, model performance was evaluated primarily using overlap-based segmentation metrics and qualitative visual assessment. While these measures provide important information regarding segmentation accuracy, future studies could incorporate additional clinically oriented evaluation criteria, including boundary-distance metrics and radiologist-based assessments, to further characterize the practical utility of the generated segmentations.

Despite these limitations, the proposed DSU-Net framework demonstrated strong quantitative and qualitative performance, providing a solid foundation for future investigations involving larger and more diverse mammography datasets.

## Conclusion

This study presented DSU-Net, a deep learning framework for automated breast lesion segmentation in mammographic images using the CBIS-DDSM dataset. The proposed approach combined encoder-decoder feature extraction with a composite optimization strategy designed to address the challenges of lesion heterogeneity, ambiguous boundaries, and severe foreground-background imbalance commonly encountered in mammography. The resulting model achieved high segmentation accuracy while maintaining strong sensitivity and specificity, indicating reliable identification of lesion regions across diverse lesion appearances.

Beyond the reported performance, the findings demonstrate that accurate lesion delineation can be achieved without requiring extensive post-processing or complex multi-stage pipelines. The ability of DSU-Net to preserve lesion morphology and boundary information is particularly important for downstream clinical tasks such as lesion characterization, radiomic feature extraction, treatment planning, and computer-aided diagnosis. By producing consistent and anatomically meaningful segmentations, the proposed framework provides a foundation for reducing manual annotation effort and improving the reproducibility of quantitative breast image analysis.

Future work will focus on evaluating the generalizability of DSU-Net across external mammography datasets acquired from different institutions and imaging systems. In addition, the integration of attention mechanisms, transformer-based feature representations, and multi-modal clinical information may further enhance segmentation robustness and support more comprehensive breast cancer assessment frameworks.

Overall, the proposed DSU-Net architecture represents a practical and effective solution for automated mammographic lesion segmentation and highlights the potential of deep learning to advance computer-assisted breast cancer detection and analysis.

**Conflict of Interest**

The authors declare that they have no conflict of interest.

**Data Availability**

The CBIS-DDSM dataset used in this study is publicly available and can be accessed through The Cancer Imaging Archive (TCIA).

**Funding**

This research received no external funding.

**Ethical Approval**

This study utilized a publicly available, de-identified dataset and did not involve direct participation of human subjects. Therefore, ethical approval was not required.

**Acknowledgments**

The authors acknowledge The Cancer Imaging Archive (TCIA) for providing access to the CBIS-DDSM dataset used in this research.